\documentclass[twocolumn,pra,superscriptaddress,longbibliography]{revtex4-1}
\usepackage[english]{babel}
\usepackage{graphicx}
\usepackage{amssymb}
\usepackage{amsmath}
\usepackage{color}
\usepackage{comment}
\usepackage{bm}

\newcommand{\up}{\uparrow}
\newcommand{\down}{\downarrow}

\newcommand{\x}{{\bf x}}

\newcommand{\bra}[1]{\langle\left.{#1}\right|}
\newcommand{\ket}[1]{\left|{#1}\right>}

\newcommand{\nn}{\nonumber}

\newcommand{\C}{{\cal C}}

\newcommand{\NN}{{\cal N}_N}
\newcommand{\binomial}[2]{\left(\begin{array}{c}#1\\#2\end{array}\right)}
\newcommand{\sch}{Schr{\"o}dinger }

\def\lba{\left(}    \def\rba{\right)}

\newcommand{\sout}[1]{}

\begin{document}

\title{ Strong-coupling ansatz for the one-dimensional Fermi gas in a
  harmonic potential }

\author{Jesper Levinsen}\email{jesper.levinsen@monash.edu}
\affiliation{Aarhus Institute of Advanced Studies, Aarhus University, DK-8000 Aarhus C, Denmark.}
\affiliation{School of Physics and Astronomy, Monash University,
  Victoria 3800, Australia.}

\author{Pietro Massignan}
\affiliation{ICFO - The Institute of Photonic Sciences - 08860 Castelldefels (Barcelona), Spain.}

\author{Georg M. Bruun} \affiliation{Department of Physics and
  Astronomy, Aarhus University, DK-8000 Aarhus C, Denmark.}

\author{Meera M. Parish}
\affiliation{School of Physics and Astronomy, Monash University, Victoria 3800, Australia.}
\affiliation{London Centre for Nanotechnology, Gordon Street, London, WC1H 0AH, United Kingdom.}

\date{\today}

\begin{abstract}
  A major challenge in modern physics is to accurately describe
  strongly interacting quantum many-body systems. One-dimensional
  systems provide fundamental insights since they are often amenable
  to exact methods. However, no exact solution is known for the
  experimentally relevant case of external confinement.  Here, we
  propose a powerful ansatz for the one-dimensional Fermi gas in a
  harmonic potential near the limit of infinite short-range
  repulsion. For the case of a single impurity in a Fermi sea, we show
  that our ansatz is indistinguishable from numerically exact results
  in both the few- and many-body limits. We furthermore derive an
  effective Heisenberg spin-chain model corresponding to our ansatz,
  valid for any spin-mixture, within which we obtain the impurity
  eigenstates analytically. In particular, the classical Pascal's
  triangle emerges in the expression for the ground-state
  wavefunction.  As well as providing an important benchmark for
  strongly correlated physics, our results are relevant for emerging
  quantum technologies, where a precise knowledge of one-dimensional
  quantum states is paramount.
\end{abstract}


\maketitle
One-dimensional systems occupy a unique place in strongly correlated
many-body physics, as many are exactly solvable via methods such as
the Bethe Ansatz.  Likewise, the exactly solvable harmonic oscillator
plays a central role in quantum mechanics. However, when one combines
these two fundamental models and considers interacting fermions in a
1D harmonic potential, there is no known solution in general
\cite{Guan2013}.  While the problem may be solved analytically for 2
particles~\cite{Busch1998}, and numerically up to $\sim$10
particles~\cite{Gharashi2013,Volosniev2013,Deuretzbacher2014,Bugnion2013,Sowinski2013,Astrakharchik2013},
the calculation rapidly becomes untenable beyond that.  Recent
theoretical works have proposed analytic forms of the lowest energy
wavefunctions near the Tonks-Girardeau (TG) limit of infinitely strong
contact interactions~\cite{Guan2009,Girardeau2010,Cui2014}, but these
do not match exact numerical
studies~\cite{Gharashi2013,Volosniev2013,Deuretzbacher2014} once the
particle number exceeds three.  Here, we present a novel, highly
accurate ansatz for the wavefunction of the two-component 1D Fermi gas
in a harmonic potential with strong repulsive interactions.

Harmonically confined 1D systems have received a considerable amount
of interest, particularly since the experimental realization in
ultracold atomic Bose~\cite{Paredes2004,Kinoshita2004} and Fermi
gases~\cite{Serwane2011,Zurn2012,Wenz2013,JochimNew,Murmann2015}.
Experimentalists have trapped fermionic $^6$Li atoms in a 1D
waveguide, with a high degree of control over both the number of
particles in two hyperfine states and the interspecies interaction
strength. This allows one to study the evolution from few to many
particles, as well as the possibility of magnetic transitions as the
interactions are tuned through the TG limit.  The approach we propose
here provides a way to tackle the regime near the TG limit, which has
been investigated in a recent experiment~\cite{Zurn2012}. In this
case, the ground-state manifold in the confined system consists of
${{N_\up+N_\down}\choose{N_\down}}$ nearly degenerate states, where
$N_\up$ ($N_\down$) is the number of spin-$\up$ (spin-$\down$)
particles.  For the ``impurity'' problem consisting of a single
$\down$ particle ($N_\down=1$) in a sea of $N_\uparrow$ majority
particles, we generate all these states in an essentially
combinatorial manner. We show that the overlap between our ansatz
wavefunction and exact states obtained by numerical calculations
exceeds 0.9997 for $N_\up\leq8$.  In particular, the overlap with the
exact ground state for large repulsive interactions is found to
extrapolate to a value $\sim0.9999$ as $N_\uparrow\to\infty$. This
remarkable accuracy shows that our ansatz effectively \emph{solves}
the strongly interacting single $\down$ problem in a harmonic
potential, from the few- to the many-body limit.

We have furthermore mapped the strongly interacting 1D problem onto an
effective Heisenberg spin chain of finite
length~\cite{Matveev2004,Matveev2008,Deuretzbacher2014}, and we derive
an analytical expression for the Hamiltonian within which our ansatz
is exact.  In this case, our ansatz for the impurity wavefunctions in
the spin basis corresponds to discrete Chebyshev polynomials.  In
accordance with the orthogonality catastrophe \cite{Anderson1967}, we
find that the overlap with the non-interacting many-body ground state
(i.e., the quasiparticle residue) tends to zero in the thermodynamic
limit $N_\up\to\infty$. However, surprisingly, in this limit the
ground-state probability distribution of the impurity is a Gaussian
only slightly broadened compared with the non-interacting ground
state.  As we argue, our effective spin model is expected to
accurately describe \emph{any} $N_\down$, $N_\up$, opening up the
possibility of addressing the strongly interacting 1D Fermi gas with
powerful numerical methods for lattice systems.  We also discuss how
our results can be extended to higher excited states and how they may
be probed in cold-atom experiments.  Our findings are of fundamental
relevance for emerging quantum technologies where an accurate
knowledge of 1D quantum states is needed, such as for engineering
efficient state transfer~\cite{state_transfer}, and for understanding
relaxation and thermalisation in out-of-equilibrium quantum systems
\cite{Eisert2015}.

\section*{Results}

\paragraph*{{\bf The model.}}
We consider $N_\uparrow$ fermions in spin state $\uparrow$ and
$N_\downarrow$ fermions in spin state $\downarrow$, both with mass
$m$, confined in a 1D harmonic potential. The total number of
particles is written as $N_\up + N_\down = N+1$, which is convenient
when we consider the impurity problem below. The Hamiltonian is thus
\begin{align}
  {\cal H}=\sum_{i=0}^N\left[-\frac{\hbar^2}{2m}\frac{\partial^2}{\partial
      x_i^2}+\frac12m\omega^2x_i^2\right]+g\sum_{i< j}\delta(x_i-x_j),
\label{eq:H}
\end{align}
where the coupling $g$ quantifies the strength of the short-range
interactions and $\omega$ is the harmonic oscillator frequency.  Note
that particles with the same spin do not interact since their
wavefunction vanishes when $|x_i - x_j| \to 0$ due to antisymmetry
under particle exchange.  Since ${\cal H}$ commutes with the total
spin operator, the eigenstates  have well defined spin projection
$S_z = (N_\up - N_\down)/2$ and total spin $S$.  In the following, we
use harmonic oscillator units where $\omega = m = \hbar = 1$.

In the TG limit, the coupling strength $g\to\infty$ and the system
simplifies significantly due to the form of the boundary conditions
when two particles approach each other. Specifically, for a given
wavefunction $\psi(\x)$, the infinite repulsion requires
$\lim_{x_{ij}\to0}\psi(\x)=0$, with $x_{ij}\equiv x_i-x_j$ the
relative coordinate for any pair of fermions with opposite spin, and
$\x=(x_0,x_1,\dots,x_N)$.  Identical fermions always obey this
condition, as mentioned above.  Since all particles experience the
same boundary conditions, the ground-state manifold for a system with
fixed $S_z$ contains ${N_\up + N_\down}\choose{N_\down}$ degenerate
states, corresponding to the number of unique configurations of $\up$
and $\down$ particles.

The simplest eigenstate of the Hamiltonian \eqref{eq:H} is the fully
ferromagnetic state, corresponding to the maximum total spin
$S = (N+1)/2$.  In this case, the spin part of the wavefunction is
always symmetric, regardless of $S_z$, and thus the wavefunction in
real space must be antisymmetric. Specifically, the wavefunction takes
the form of a Slater determinant of single-particle harmonic
oscillator wavefunctions, and can be written \cite{Girardeau2001}
\begin{align}
  \psi_0(\x) &=\NN\left(\prod_{0\leq i<j\leq
      N}x_{ij}\right)e^{-\sum_{k=0}^Nx_k^2/2} ,
\label{eq:psi0}
\end{align}
with normalization
\begin{align}\notag
  \NN=\frac1{\sqrt{(N+1)!}}\sqrt{\frac{2^{\frac12N(N+1)}}
    {\pi^{\frac12(N+1)}\prod_{n=0}^Nn!}} \ .
\end{align}
The ferromagnetic state corresponds to that of $N+1$ identical
fermions, and thus its energy is
$E_0 = \sum_{n=0}^{N} (n+1/2)-1/2 = N(N+2)/2$, where we subtract the
center-of-mass zero point energy.  Furthermore, it is an eigenstate
for all $g$ since the wavefunction antisymmetry guarantees that it
vanishes when $x_{ij} \to 0$ so that it does not experience the
particle-particle interaction.

\begin{centering}
\begin{figure}
\vskip 0 pt
\includegraphics[width=0.95\columnwidth,clip]{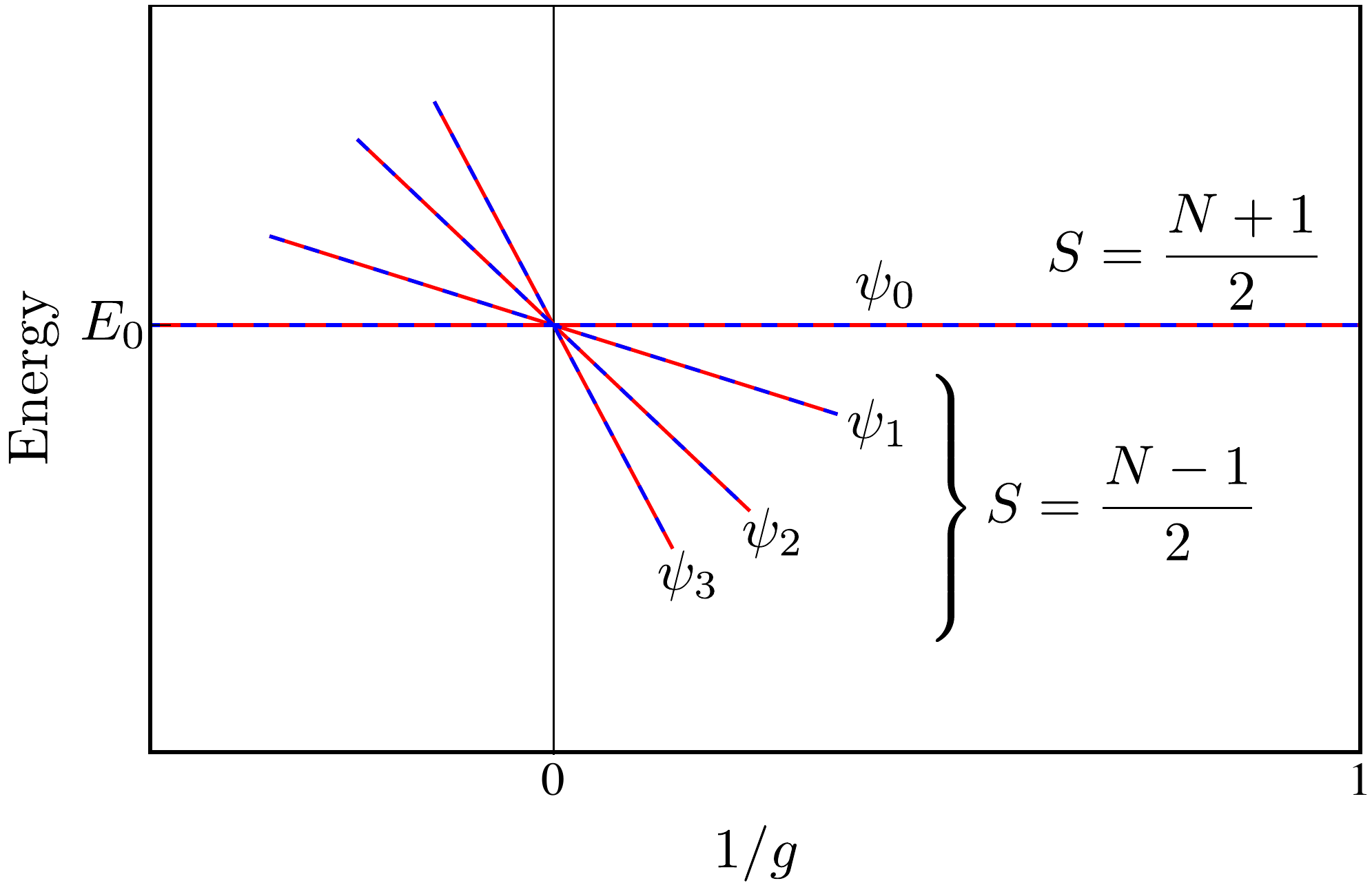}
\caption{ {\bf Energy levels in the Tonks-Girardeau limit.}  We
  display the exact energies $\langle\psi_l|{\cal H}|\psi_l\rangle$
  (red) and the result of our ansatz
  $\left<\tilde\psi_l\right|{\cal H}\left|\tilde\psi_l\right>$ (blue
  dashed), given by $E_0 - {\mathcal C}_l/g$ in this limit, for the
  case of one $\down$ particle and $N=3$ $\up$ particles.  For $g<0$,
  there is also a two-body bound state at negative energies which we
  do not show.}
\label{fig:energy}
\end{figure}
\end{centering}

For a given $S_z$ (corresponding to fixed $N_\uparrow$ and
$N_\downarrow$), the remaining eigenstates with the same energy $E_0$
in the TG limit may be characterized by other values of $S$. For
instance, for $N_\up = N_\down=1$, there are two states, characterized
by either $S=0$ or $S=1$. However, for general particle number, spin
alone is not sufficient to determine the states with $S < (N+1)/2$,
since the degeneracy of the ground-state manifold is
${N_\up+N_\down}\choose{N_\down}$ whereas the number of different $S$
for a given $S_z$ is $1+{\rm min}(N_\up,N_\down)$.  Thus, in order to
construct a unique orthogonal basis of eigenstates in the TG limit, we
must consider how the states in the ground-state manifold evolve as
$g\to \infty$, as illustrated in Fig.\ \ref{fig:energy}. We mostly
focus on the impurity problem where we have one $\down$ particle at
position $x_0$ and $N$ $\up$ particles at positions $x_i$ with
$1\leq i\leq N$. In this case, we have $N$ eigenstates with spin
$S=S_z=(N-1)/2$, in addition to the ferromagnetic state.
\\

\paragraph*{\bf Ground-state manifold in the TG limit.}
To construct the wavefunctions for the impurity problem with
$S=S_z=\frac{N-1}2$, it is useful to define a complete (but not
orthogonal) set of basis functions involving $\phi_0=\psi_0(\x)$ and
the $N$ states:
\begin{align}
  \phi_l&=\psi_0(\x)\sum_{1\leq i_1<\cdots<i_l\leq N}s_{i_1}\cdots
  s_{i_l}, \hspace{5mm} 1\leq l\leq N,
\label{eq:basis}
\end{align}
where $s_{i}\equiv \mbox{sign}(x_{i0})$. For simplicity, we omit the
dependence of $\phi_l$ on the coordinates. Each sign function simply
replaces a zero-crossing in the Slater determinant (\ref{eq:psi0})
with a cusp at the position where the impurity meets a majority
($\uparrow$) particle ($x_{i0}=0$).  As an example, for $N=2$ we have
basis functions:
\begin{align} \nn
\phi_0 = & \ {\cal N}_2 \ x_{12} x_{01} x_{02} \ e^{-\left(x_0^2+x_1^2+x_2^2\right)/2}, \\ \nn
\phi_1 = & \ {\cal N}_2 \   x_{12}  \left(|x_{01}| x_{02} + x_{01} |x_{02}|  \right) \ e^{-\left(x_0^2+x_1^2+x_2^2\right)/2}, \\ \nn
\phi_2 =  & \ {\cal N}_2 \ x_{12} |x_{01} x_{02}| \ e^{-\left(x_0^2+x_1^2+x_2^2\right)/2}.
\end{align}
The basis functions are clearly degenerate with the ferromagnetic
state \eqref{eq:psi0} when $g\rightarrow\infty$, since the interaction
energy vanishes while the energy of motion in the harmonic potential
is the same for all $\phi_l$. This can be shown by noting that for
any ordering of the particles (say $x_0 < x_1 < ... < x_N$) we have
$\phi_l \propto \psi_0(\x)$. Thus all eigenstates of the ground state
manifold in the TG limit must be linear combinations of the basis
functions. Note that alternatively we could have chosen a basis set
whose functions are non-zero only for a particular ordering of
particles as in Ref.~\cite{Deuretzbacher2008}.

The central question we address here concerns the nature of the
eigenstates in the vicinity of the TG limit, i.e., we wish to know the
wavefunctions and energies perturbatively in the small parameter
$1/g$. This allows one to uniquely define the eigenstates at
$g\rightarrow\infty$ as being those that are adiabatically connected
to the states at finite $g$.  Before proceeding with degenerate
perturbation theory, it is instructive to consider the structure of
the exact eigenstates $\psi_l$ (up to corrections of order $1/g$) for
$N=1$ \cite{Busch1998} and $N=2$ \cite{Guan2009},
\begin{align} \nn
  N=1: \hspace{3mm} & \psi_0=\phi_0, \hspace {2mm} \psi_1=\phi_1, \\
  N=2: \hspace{3mm} & \psi_0=\phi_0, \hspace {2mm}
  \psi_1=\sqrt{\frac38}\phi_1, \hspace{2mm}
  \psi_2=\sqrt {\frac1 8}(\phi_0-3\phi_2).
\label{eq:n1andn2}
\end{align}
The subscripts on the wavefunctions order these in terms of decreasing
energy for small but positive $1/g$.  Note that the eigenstates split
into two orthogonal sets which are even or odd with respect to parity,
since the Hamiltonian commutes with the parity operator.  Referring to
Fig.~\ref{fig:energy} and focussing on the repulsive case $g>0$, we
see that the ferromagnetic state $\psi_0$ has the maximum energy
within the manifold \cite{Cui2014}, while the ground state $\psi_N$
has the lowest total spin, i.e., $S=S_z$, in accordance with the
Lieb-Mattis theorem~\cite{Lieb1962}. Physically, the cusps in the
wavefunction for $g\rightarrow \infty$ can easily be shifted from
zero, which decreases the kinetic energy and thus leads to a lower
energy compared with the ferromagnetic state. Indeed we see two
patterns emerging: $\psi_1$ contains only states with one cusp, and
only the ground state $\psi_N$ contains the state with the maximal
number of cusps. These observations suggest that the system may lower
its energy by successively acquiring more cusps in the wavefunction.

Inspired by the above considerations, we now propose the following
strong-coupling ansatz for the impurity eigenstates of the
ground-state manifold in the vicinity of the TG limit:
\begin{itemize}
\item {\em For any $N$, the exact wavefunction $\psi_l$ essentially
    corresponds to $\tilde\psi_l$, a superposition of the basis
    functions $\phi_k$ restricted to $k\leq l$.}
\end{itemize}
In other words, the wavefunctions are obtained by a Gram-Schmidt
orthogonalization scheme on the set of basis functions $\{\phi_k\}$:
$\tilde\psi_1$ is obtained by adding one cusp to $\psi_0$,
$\tilde\psi_2$ is obtained by adding one more cusp and then
orthogonalising it to $\tilde\psi_0$, and so on.  We emphasize that
the ansatz allows one to obtain the entire ground-state manifold using
linear algebra manipulations only.  Thus, one can go far beyond the
limit $N\lesssim9$ of current state of the art calculations
\cite{Volosniev2013}. In fact, we will see that the ansatz allows us
to obtain \emph{analytic} expressions for all wavefunctions in the
ground-state manifold for any $N$.  We will show that the ansatz is
remarkably accurate compared with exact numerical results, and that it
allows one to calculate several observables analytically, even in the
many-body limit.

The procedure for constructing our ansatz wavefunctions $\tilde \psi$
as outlined above can in fact be performed straightforwardly even for
large $N$, by noting that the inner products of the basis functions
\eqref{eq:basis}, $\Phi_{ln}\equiv\langle \phi_l|\phi_n\rangle$, may
be calculated combinatorially (see Methods).
\\

\paragraph*{\bf Perturbation theory around the TG limit.}
To demonstrate the accuracy of our ansatz, we now turn to the explicit
solution of the \sch equation in the vicinity of the TG limit.  Since
here there are $N+1$ degenerate states, we apply degenerate
perturbation theory and obtain the ground-state manifold by means of
finite-matrix diagonalization, in a similar manner to
Refs.~\cite{Deuretzbacher2014,Volosniev2013}.  The energy can be
written as $E \simeq E_0 -\mathcal{C}/g$, where $\mathcal{C}$ is the
1D contact density \cite{Olshanii2003,Tan2008,Barth2011}. From the
Hellmann-Feynman theorem, we then obtain
\begin{align} \nn
  \mathcal{C} = - \left.\frac{dE}{d (g^{-1})}\right|_{g \to \infty} =
  - \left. \left< \frac{\partial {\cal H}}{\partial (g^{-1})} \right>
  \right|_{g \to \infty} \equiv \frac{\bra{\Psi} {\cal H}'
    \ket{\Psi}}{\left<\Psi | \Psi \right>},
\end{align}
which defines the perturbation ${\cal H}'$ due to a non-zero
$1/g$. The state $|\Psi\rangle$ is a linear combination of the basis
states $\{\phi_l\}$ of the ground-state manifold:
$\ket{\Psi} = \sum_n \alpha_{n} \ket{\phi_n}$.  To obtain the
eigenstates, we require $\ket{\Psi}$ to be a stationary state,
i.e.~$\frac{\delta \mathcal{C}}{\delta \alpha_l^*} = 0$, resulting in
the matrix equation $\Phi^{-1} {\cal H}' \alpha = \mathcal{C} \alpha$,
with ${\mathcal C}$ the eigenvalue (contact density) of the state
$|\Psi\rangle$. The matrix elements of ${\cal H}'$ are
\begin{align} {\cal H}'_{ln} & 
= \sum_{i=1}^N \int d\x \ \delta(x_{i0}) \left. \frac{\partial
      \phi_l }{\partial x_{i0}} \right|^{x_{i0}=0^+}_{x_{i0}=0^-} \left. \frac{\partial
      \phi_n }{\partial x_{i0}} \right|^{x_{i0}=0^+}_{x_{i0}=0^-},
\label{eq:Hp}
\end{align}
(see Methods).

For $N=1$ ($N=2$) the evaluation of ${\cal H'}$ is straightforward and
yields ${\cal H'}_{11}=2\sqrt{2/\pi}$
(${\cal H}'_{11}={\cal H}'_{22}=9/\sqrt{2\pi}$), while all other
elements vanish. Thus we find
\begin{align}
  N=1: & \hspace{5mm}
  \left(\begin{array}{c}\C_0 \\ \C_1
    \end{array}\right)=\sqrt{\frac8\pi}\left(\begin{array}{c} 0 \\ 1
    \end{array}\right), 
\label{eq:C1}\\
  N=2: & \hspace{5mm}
  \left(\begin{array}{c}\C_0 \\ \C_1 \\
      \C_2\end{array}\right)=\frac{27}{8\sqrt{2\pi}}\left(\begin{array}{c} 0 \\ 1 \\ 3
    \end{array}\right),
\label{eq:C2}
\end{align}
where $\C_l\equiv\langle\psi_l|{\cal H}'|\psi_l\rangle$ is the contact
coefficient corresponding to the state $\psi_l$. All the eigenstates
for $N\leq2$ are in fact uniquely determined by the two symmetries of
parity and spin, so that the ratios of $\C_l$ and the general
structure of the wavefunctions in Eq.~\eqref{eq:n1andn2} hold for any
confining potential that preserves parity and spin.  However, these
symmetries alone are not sufficient to determine the eigenstates for
$N>2$, and therefore $N=3$ will provide a non-trivial test of our
ansatz. In this case, the coefficients of ${\cal H}'$ and the
eigenstates may still be evaluated analytically, but their form is
sufficiently complicated that we relegate these to the Methods.
Converting long analytical expressions into numerical values for
brevity, we obtain
\begin{align}
N=3:\hspace{5mm} &
\psi_0=\phi_0, \nn \\
& \psi_1=\sqrt{\frac15}(1.00188\phi_1
-0.00941 %221
\phi_3),\nn \\
&  \psi_2=\frac12(\phi_0-\phi_2),\nn \\
&\psi_3=\sqrt{\frac1{20}}(0.99246 
\phi_1-4.99996\phi_3).
\label{eq:psi3}
\end{align}
while the contact coefficients are
\begin{align}
  N=3: & \hspace{5mm}
  \left(\begin{array}{c}\C_0 \\ \C_1 \\
      \C_2 \\ \C_3 \end{array}\right)=1.18067\left(\begin{array}{c} 0 \\ 1.00305 \\
      3.02818 \\ 6
    \end{array}\right).
\label{eq:C3}
\end{align}
These contact coefficients determine the energy splitting shown in
Fig.~\ref{fig:energy} and they agree with those obtained in
Ref.~\cite{Deuretzbacher2014}.  For $N\geq4$ we resort to a numerical
evaluation of the matrix elements of ${\cal H}'$ which may be
calculated efficiently using a novel method outlined in the
Methods.

\begin{centering}
\begin{figure}
\vskip 0 pt
\includegraphics[width=\columnwidth,clip]{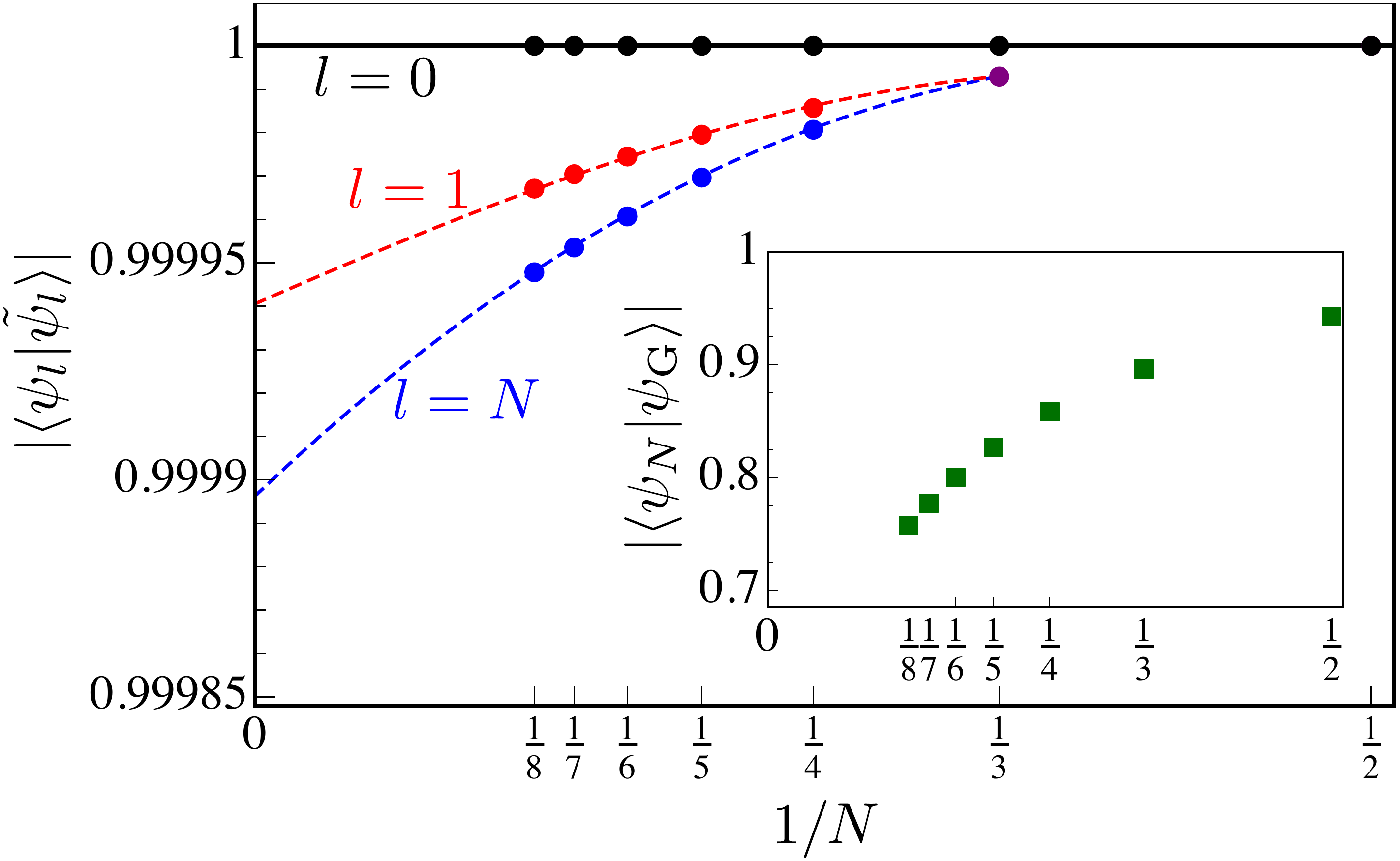}
\caption{ {\bf Accuracy of the ansatz.}  Overlaps between our ansatz
  $\tilde \psi_l$ and the exact wavefunctions $\psi_l$ for majority
  particle numbers $N\leq8$. For the ferromagnetic state, this always
  equals 1 (black line). The red and blue dots depict the overlap for
  $\tilde\psi_1$ and $\tilde\psi_N$ (ground state) respectively.
  These are both 1 for $N=2$, as these states are uniquely determined
  by spin and parity, while they are both 0.999993 for $N=3$.  The
  extrapolations (dashed lines) are least-squares fits of the data
  points to cubic polynomials. Inset: The wavefunction overlap of
  Girardeau's proposed state \cite{Girardeau2010} with the exact
  ground state.}
\label{fig:fidelity}
\end{figure}
\end{centering}

Our ansatz, however, is far simpler than the brute-force approach
above. While the evaluation of the multi-dimensional integral in
Eq.~\eqref{eq:Hp} quickly becomes untenable as $N$ increases above
$\sim10$, the implementation of our ansatz is a basic exercise in
linear algebra: Applying our Gram-Schmidt orthogonalization scheme, we
find the states
\begin{align}
  N=3: \hspace{5mm} &\tilde\psi_0=\phi_0, \hspace {2mm}
  \tilde\psi_1=\sqrt{\frac15}\phi_1, \hspace{2mm}
  \tilde\psi_2=\frac12(\phi_0-\phi_2),\nn
  \\
  & \tilde\psi_3=\sqrt{\frac1{20}}(\phi_1-5\phi_3).
\label{eq:blumeresult}
\end{align}
Comparing Eq.~\eqref{eq:blumeresult} with Eq.~\eqref{eq:psi3}, we see
that our ansatz is extremely accurate, with only a minute deviation
from the exact result for $\tilde\psi_1$ and $\tilde\psi_3$ ($\psi_0$
and $\psi_2$ are determined exactly from parity and spin).  We note
that our proposed wavefunctions are identical to those obtained
numerically in Ref.~\cite{Gharashi2013}, illustrating that the results
of our ansatz are essentially indistinguishable from numerical
calculations.

We now demonstrate explicitly that the very high accuracy of our
ansatz holds also for higher particle number $N$, and that it even
seems to hold in the many-body limit.  A natural measure of its
accuracy is the wavefunction overlap
$|\bra{\psi_l}\tilde\psi_l\rangle|$ between the exact eigenstates
$\psi_l$ and our proposed ones $\tilde\psi_l$. Writing the
wavefunctions as $\psi_l=\sum_{n=0}^NL_{ln}\phi_n$ and
$\tilde\psi_l=\sum_{n=0}^N\tilde L_{ln}\phi_n$, the overlap is simply
$|(\tilde L \Phi L^T)_{ll}|$. For the two non-exact states with $N=3$
discussed above, we then find this quantity to be 0.999993, where we
remind the reader that this is the numerical value of an analytic
result (see Methods).  Strikingly, we find that the overlap exceeds
$0.9997$ for all states up to $N=8$, with the error being largest for
the states ``intermediate'' between the ferromagnetic state $\psi_0$
and the ground state $\psi_N$.  In Fig.~\ref{fig:fidelity}, we
illustrate how the wavefunction overlaps for the $\tilde\psi_1$ and
$\tilde\psi_N$ always exceed 0.99994. In addition, the overlap in the
ground state appears to extrapolate to a value $\sim0.9999$ as
$N\to\infty$.  Our ansatz is therefore essentially indistinguishable
from ``numerically exact'' methods, even in the many-body limit.  This
shows that our ansatz effectively \emph{solves} the strongly
interacting 1D impurity problem for general $N$.

Of particular interest is the state $\psi_N$, the ground state for
repulsive interactions. Girardeau proposed \cite{Girardeau2010} that
this state is simply given by the state with the maximum number of
cusps inserted, i.e., $\psi_{\rm \tiny G}=\phi_N$. As shown in the
inset of Fig.~\ref{fig:fidelity}, the overlap of Girardeau's proposed
state with the exact ground state is $76\%$ for $N=8$, and it most
likely tends to zero as $N\to\infty$.  Thus, our ansatz is a
significant improvement compared to previous proposals for the
ground-state wavefunction.

\begin{centering}
\begin{figure}
\vskip 0 pt
\includegraphics[width=\columnwidth,clip]
{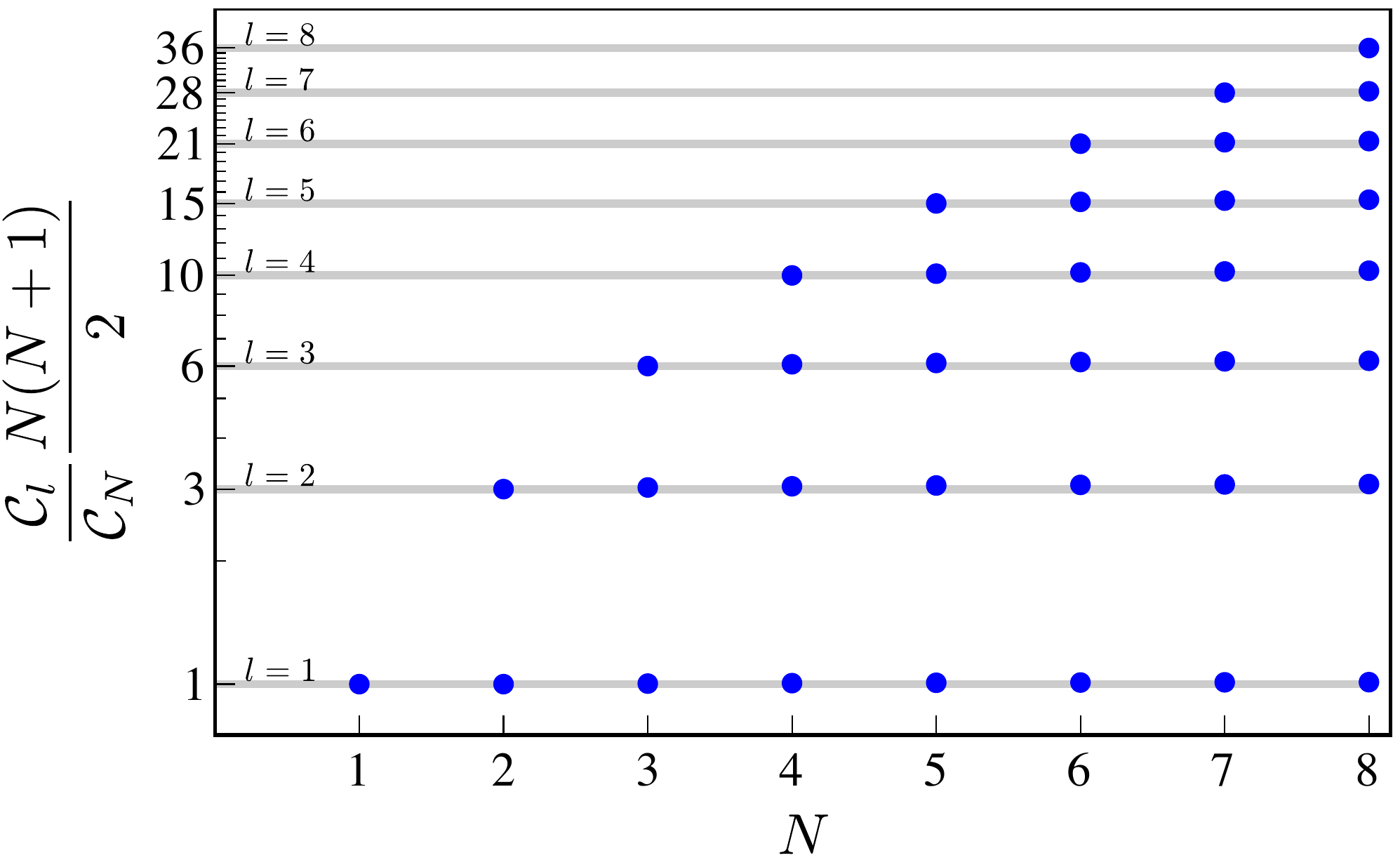}
\caption{ {\bf Contact coefficients of the ground state manifold.}
  The contact coefficients (blue dots) of the exact eigenstates
  $\psi_l$ in the Tonks-Girardeau regime, controlling the splitting of
  the energy levels at finite but large coupling. The gray lines
  represent the approximate relationship
  $\frac{\C_l}{\C_N}=\frac{l(l+1)}{N(N+1)}$, see
  Eq.~\eqref{eq:spectrum}.}
\label{fig:contact}
\end{figure}
\end{centering}

We now turn to the contact coefficients of the $N+1$ energy levels in
the ground-state manifold, i.e., the splitting of the spectrum at
finite coupling. In Fig.~\ref{fig:contact} we show how the energy
takes the following approximate form
\begin{align}
E_l\simeq E_0-\frac{\C_N}g\frac{l(l+1)}{N(N+1)}.
\label{eq:spectrum}
\end{align}
Comparing with Eqs.~(\ref{eq:C1}), \eqref{eq:C2} and \eqref{eq:C3}, we
see that this expression is exact for $N=1$ and $2$, while it holds to
within $3.0\%$ for $N\leq8$. We show in the next section that the
spectrum given by Eq.~\eqref{eq:spectrum} is intimately linked with an
effective Heisenberg spin Hamiltonian within which our ansatz is
exact.
\\

\paragraph*{\bf Effective Heisenberg spin chain.}
We now discuss how the 1D problem can be mapped onto a Heisenberg spin
model~\cite{Matveev2004,Matveev2008,Deuretzbacher2014}. This enables
us to determine the states $\tilde \psi_l$ analytically, and it also
allows us to generalise our ansatz for the impurity problem to
\emph{any} $N_\downarrow$.

In the limit $g\to\infty$, the system consists of impenetrable
particles since the wavefunction must vanish when two particles
approach each other.  Thus, if the particles are placed in a
particular order, they should retain that ordering as long as the
repulsion is infinite.  This allows us to consider the system in the
TG limit as a discrete lattice of finite length $N+1$, where the
particle furthest to the left is at site $i=0$, the next particle is
at site $i=1$ and so on.  A small but finite value of $1/g$ then
allows neighboring particles to exchange position, introducing a
nearest-neighbor spin interaction in the lattice picture. We can thus
write the Hamiltonian in the lattice as
\begin{align} {\cal H} \simeq E_0 - \frac{{\cal
      H}'}{g}=E_0+\frac{\C_N}g\sum_{i=0}^{N-1} \left[ J_i\hat{\bf
      S}^{i}\cdot\hat{\bf S}^{i+1}-\frac14 J_i\right],
\label{eq:lattice}
\end{align}
where $\hat{\bf S}^{i}$ is the spin operator at site $i$ and $J_i$ is the
nearest neighbor exchange constant, which can in general depend on
$i$~\cite{Volosniev2013,Volosniev2014}.  Subtracting the constant in
each term of the sum ensures that the ferromagnetic state has energy
$E_0$.  The Hamiltonian \eqref{eq:lattice} is valid to linear order in
$1/g$ and the general form holds for any external potential.

The couplings $J_i$ in the Heisenberg model (\ref{eq:lattice}) can be
determined by considering the single $\down$ impurity problem in a new
basis of position states $\ket{\down_i}$ with $0\leq i\leq N$. The
lattice position $i$ corresponds to the position of the impurity
relative to the $N$ majority particles.  The position states are
orthonormal, $\bra{\down_i}\left. \down_j\right>=\delta_{ij}$, and can
be related back to $\phi_l$ (see Methods). The perturbation
${\cal H}'$ may then be evaluated in the position basis of the
impurity by inserting a complete set of eigenstates, yielding
\begin{align}
\bra{\down_i}{\cal H}'\ket{\down_j}= \sum_{l=0}^N
\bra{\down_i}\psi_l\rangle \,
\mathcal{C}_l \,
\langle\psi_l \ket{\down_j }.
\label{eq:Hij}
\end{align}
The matrix elements (\ref{eq:Hij}) provide an explicit construction of
the Heisenberg Hamiltonian (\ref{eq:lattice}).

\begin{centering}
\begin{figure}
\vskip 0 pt
\includegraphics[width=\columnwidth,clip]{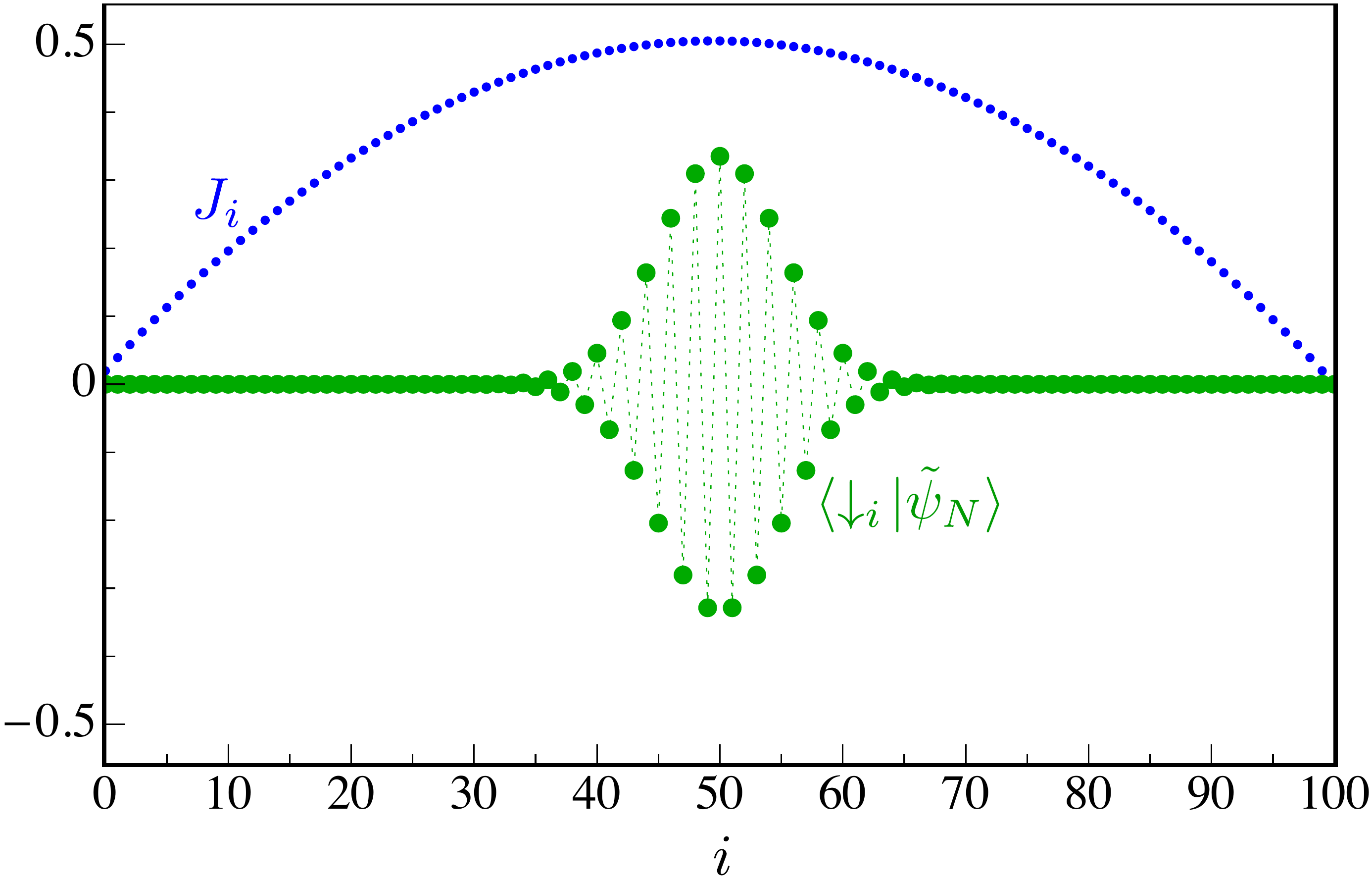}
\caption{
{\bf Exchange constants and ground state of the Heisenberg model.}
Illustration of the nearest-neighbor exchange constants
  \eqref{eq:J} of the spin Hamiltonian \eqref{eq:lattice} for $N=100$.
  We also show the ground-state wavefunction \eqref{eq:GS} within our
  ansatz (green dots).}
\label{fig:lattice}
\end{figure}
\end{centering}

We now determine the Heisenberg Hamiltonian within which our
strong-coupling ansatz for the eigenstates is exact.  Proceeding via
``reverse engineering'', we form the effective Hamiltonian by
replacing $\psi_l$ with our ansatz wavefunctions $\tilde\psi_l$ in
Eq.~(\ref{eq:Hij}).  By inspection of the effective Hamiltonian for
all $N\leq100$, we find that we must use the approximation
$\C_l \simeq \C_N \frac{l(l+1)}{N(N+1)}$ from Eq.~\eqref{eq:spectrum}
in Eq.~(\ref{eq:Hij}) in order to obtain a Hamiltonian restricted to
nearest-neighbor interactions, and we obtain the couplings
\begin{align}
  J_i&=\frac{-\left(i-\frac{N-1}2\right)^2+\frac14(N+1)^2}{N(N+1)/2}.
\label{eq:J}
\end{align}
The exchange constant takes the form of an inverted parabola and is
thus reminiscent of the real space harmonic oscillator potential (see
Fig.~\ref{fig:lattice}).  The form of the coefficients means that the
impurity at small positive $1/g$ may minimize its energy by occupying
primarily the center of the spin chain, while alternating the sign of
the wavefunction on the different sites.  Contrast this with the
ferromagnetic state, which is a completely symmetric function of the
impurity position
$\ket{\psi_0}=\frac1{\sqrt{N+1}}\sum_{i=0}^N\ket{\down_i}$.  This is
equivalent to the state obtained by applying the total spin lowering
operator $\hat{S}_- = \sum_i \hat{S}^i_-$ to the spin polarized state
with $S_z = (N+1)/2$.  Note that this symmetric spin function
corresponds to an antisymmetric wavefunction in real space.

The Heisenberg model obtained from our ansatz is exact for $N=1$ and
$N=2$, while it is approximate for larger $N$. In particular, for
$N=3$ our ansatz yields
\begin{align}
  \left(\begin{array}{c}
      J_0\\J_1\\J_2\end{array}\right)=\frac12\left(\begin{array}{c}1\\4/3\\1
\end{array}\right),
\end{align}
which should be compared with the  result obtained by using the exact eigenstates and energies in Eq.~(\ref{eq:Hij}), yielding 
\begin{align} 
\label{eq:J3} \left(\begin{array}{c}
      J_0\\J_1\\J_2\end{array}\right)=\frac12\left(\begin{array}{c}
      1.009\\1.325\\1.009
\end{array}\right).
\end{align}
The error in the coefficients is thus less than $1\%$.  Note that
Eq.~\eqref{eq:J3} agrees with that of Ref.~\cite{Deuretzbacher2014}.
For larger $N\leq8$ we find that the error in the coefficients at the
central sites remains $\lesssim0.3\%$, while the error at the edges of
the spin chain remains $\lesssim5\%$. This shows that our ansatz is
most accurate when the impurity is near the center of the harmonic
potential, which is always the case for the ground-state wavefunction
as we demonstrate below.

Our effective ``harmonic'' Heisenberg model allows us to determine the
general solution for the single $\down$ impurity within our ansatz
\emph{analytically}. We obtain
\begin{align}
  \left|\tilde\psi_l\right> =&\eta_l^{(N)}\sum_{i=0}^N\sum_{n=0}^l
  (-1)^n{{l+n}\choose{n}}{{N-n}\choose{N-l}}{{i}\choose{n}}\left|\down_i\right>,
  \label{eq:Exact}
\end{align}
for the eigenstates in the ground-state manifold, where
$\eta_l^{(N)}=[{{N+l+1}\choose{2l+1}}{{2l}\choose{l}}]^{-1/2}$ is a
normalization constant. This result may be verified by direct
application of the Hamiltonian \eqref{eq:lattice}, and follows from
the basis functions $\phi_l$ being discrete polynomials of the
variable $(i-N/2)$ of maximum order $l$ in the spin chain. The
Gram-Schmidt procedure of our ansatz then yields the orthonormal
discrete polynomials $\tilde\psi_l$ with maximal order $l$ in the
variable $(i-N/2)$. The functions in Eq.~\eqref{eq:Exact} are
well-known in the field of approximation theory as discrete Chebyshev
polynomials --- see, e.g., Ref.~\cite{Beals2010}.  The analytical form
for the ansatz wavefunctions provides a simple solution to the
Gram-Schmidt procedure for general $N$.  In particular, the
ground-state wavefunction is simply a (sign-alternating) Pascal's
triangle:
\begin{align}
  \ket{\tilde \psi_N} = {{2N}\choose{N}}^{-1/2}
  \sum_{i=0}^N(-1)^i{{N}\choose{i}}\ket{\down_i}.
\label{eq:GS}
\end{align}
Note that, in real space, this wavefunction does not change sign under
the exchange of the impurity with a majority particle. 

From the analytical expression (\ref{eq:GS}), we can determine the
probability that the impurity is at position $i$ relative to the
majority particles in the ground state. We obtain
$P_N(i)=|\bra{\down_i}\psi_N\rangle|^2\simeq
|\bra{\down_i}\tilde\psi_N\rangle|^2=
{{2N}\choose{N}}^{-1}{{N}\choose{i}}^2$.
This prediction is dramatically different from the constant
probability distribution $P_G(i)=1/(N+1)$ predicted by Girardeau's
proposed ground state, which in the spin-chain model takes the form
$\ket{\psi_{\rm \tiny G}}=(N+1)^{-1/2}\sum_{i=0}^N(-1)^i\ket{i}$.
Indeed, we see that
$|\langle\tilde\psi_N|\psi_{\rm \tiny G}\rangle|^2\approx\sqrt{\pi/N}$
as $N\to\infty$.  Thus, $\psi_G$ is inaccurate for the ground state in
the harmonic potential.

Finally, we emphasize that the mapping to the effective Heisenberg
model allows us to find solutions for \emph{any} $N_\up$ and
$N_\down$: One simply needs to calculate the eigenstates of the
Hamiltonian \eqref{eq:lattice} with coefficients given by
Eq.~\eqref{eq:J}.  For instance, in the case of $N_\up=N_\down=2$, the
ground-state manifold is spanned by the 6 states
$S_-^iS_-^j\ket{\up\up\up\up}$ with $i\neq j$.  Within this basis, we
find overlaps $\gtrsim0.99998$ between exact and approximate
eigenstates, and our results are in excellent agreement with the
wavefunctions obtained numerically in Ref.~\cite{Gharashi2013}.
Furthermore, within our ansatz the contact coefficients of the six
states take the form $\C_3(0, 1, 5 - \sqrt{7}, 3, 6, 5 + \sqrt{7})/6$
where $\C_3$ is the contact coefficient from the
$(N_\up,N_\down)=(3,1)$ problem --- see Eq.~\eqref{eq:C3}. As
expected, since the Hamiltonian commutes with the spin operator, the
spectrum contains that of the single-impurity problem and, in
accordance with the Lieb-Mattis theorem \cite{Lieb1962}, the ground
state has $S=0$.
\\

\begin{centering}
\begin{figure}
\vskip 0 pt
\includegraphics[width=\columnwidth,clip]{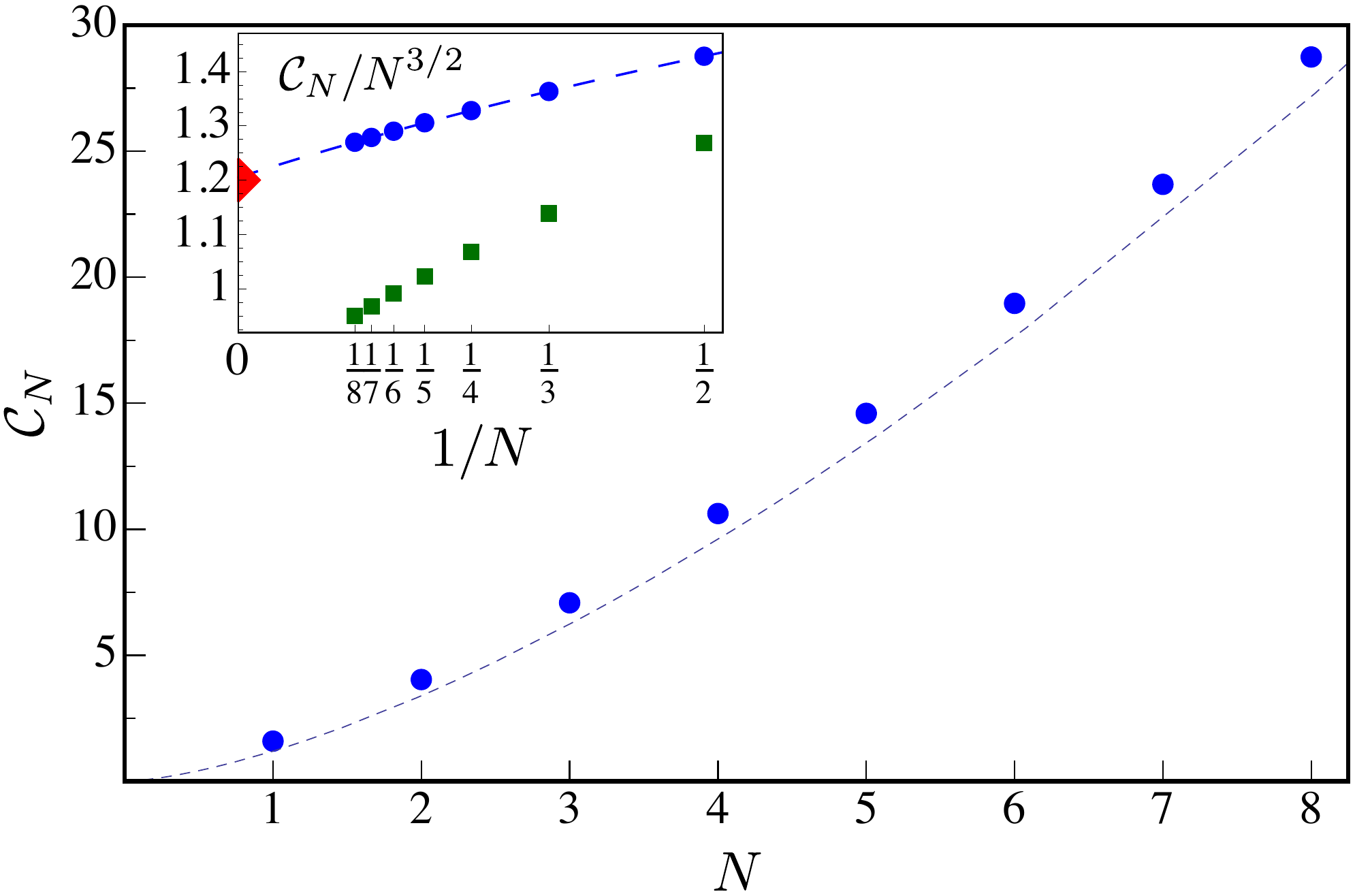}
\caption{ {\bf Contact of the ground state in the few- and many-body
    limit.}  Contact coefficient of the ground state at small positive
  $1/g$ as a function of $N$. The dots are the analytical results for
  $N\leq3$ and numerical results for $4\leq N\leq8$. We do not show a
  comparison between the ground-state contact and the perturbation
  evaluated within our approximate states,
  $\bra{\tilde\psi_N}{\cal H}'\ket{\tilde\psi_N}$, since the relative
  error between these is less than $0.05\%$ for $N\leq8$. The dashed
  line is McGuire's free-space solution mapped to the harmonic
  potential using the local density approximation --- see the
  discussion in the main text. Inset: The ground-state contact
  coefficient in units of $N^{3/2}$ and plotted as a function of $1/N$
  to illustrate the possible convergence to McGuire's prediction
  (marked by a triangle). The dashed line is a cubic fit to our
  data. We also compare with the expectation value of Girardeau's
  proposed ground state
  $\bra{\psi_{\rm G}}{\cal H}'\ket{\psi_{\rm G}}$ (green squares).}
\label{fig:contact2}
\end{figure}
\end{centering}

\paragraph*{\bf Approaching the many-body limit.} 
The fact that the wavefunction overlaps appear to extrapolate to a
numerical value very close to 1 (see Fig.~\ref{fig:fidelity}),
indicates that our ansatz is highly accurate also in the many-body
limit.  Thus, we now investigate the limit $N\rightarrow\infty$ for
the impurity ground state \eqref{eq:GS} at large repulsion.  We focus
on properties that depend on the impurity probability distribution in
the bulk of the system.

The first such quantity is the contact coefficient, as shown in
Fig.~\ref{fig:contact2}. We compare it with the expression
$\C_N\approx8\sqrt{2N^3}/(3\pi)$ corresponding to McGuire's exact
solution to the single impurity problem in free space
\cite{McGuire1965} mapped onto the harmonically confined system using
the local density approximation \cite{Astrakharchik2013}. We see that
our prediction for the contact appears to extrapolate to the Bethe
ansatz result in the many-body limit, thus implying that the local
density mapping is valid for the single-impurity ground-state energy.
Indeed, this is consistent with the fact that the impurity
ground-state wavefunction is confined to the central region of the
trap (see Fig.~\ref{fig:lattice}) where the density of majority
particles is highest.

We next calculate the probability density of the impurity in real
space, $P_N(x_0)=\int dx_1\cdots dx_N|\psi_N(\x)|^2$.  This is very
complicated to evaluate for general $N$, but in the thermodynamic
limit $N\rightarrow\infty$, the probability distribution of the
approximate ground-state wavefunction~\eqref{eq:GS} may be converted
into $P_N(x_0)$.  The distribution of majority particles is unaffected
by the presence of the impurity in the thermodynamic limit, and
according to the local density approximation it is
$n(x) = \frac{1}{\pi} \sqrt{2 \mu_0 - x^2}$, where $\mu_0$ is the
chemical potential at the center of the harmonic potential. This in
turn yields
$N = \int^{\sqrt{2\mu_0}}_{-\sqrt{2\mu_0}} n(x) dx = \mu_0$. The
lattice index $i$ in the Heisenberg model corresponds to the number of
majority particles to the left of the impurity; thus it may be related
to the position in real space via $i=\int_{-\infty}^{x_0}n(x)$. Since
$\int_{-\infty}^0 n(x)=N/2$, we can then write:
\begin{align}
  \frac{i-N/2}{N}  &= \frac{1}{N} \int_0^{x_0} n(x) dx %\nn \\  
  \simeq \frac{2x_0}{\pi\sqrt{2\mu_0}}
\end{align}
where we have taken the central part of the harmonic potential with
$x_0 \ll \sqrt{2\mu_0}$, Substituting this into Stirling's
approximation to the ground-state probability distribution,
$P_N(i)\approx 2(\pi N)^{-1/2}\exp[-(2i-N)^2/N]$, finally yields the
probability density of the impurity particle in the thermodynamic
limit:
\begin{align}
  P_N(x_0) \simeq \lba \frac{2}{\pi} \rba^{3/2} e^{-8 x_0^2/\pi^2}.
\label{eq:PNz}
\end{align}
Remarkably, upon tuning the system from the non-interacting ground
state at $g=0^+$ with probability density
$P_{\rm \tiny NI}(x_0)=\exp(-x_0^2)/\sqrt\pi$, the impurity
wavefunction has only spread out slightly in the TG limit. The broader
distribution may be viewed as an increase in the harmonic oscillator
length by a factor $\pi/2\sqrt2$ or as a decreased effective mass.
Note that our predicted probability density is completely different
from that of the Girardeau state, $n(x_0)/N$, which equals that of the
ferromagnetic state. In particular, our predicted distribution retains
a width $\sqrt{\left<x^2\right>}\sim1$ for any $N$, while for the
ferromagnetic state the width is
$\sqrt{\left<x^2\right>}\sim \sqrt{N}$.  The narrow width of the
ansatz ground state compared with the length of the spin chain in turn
implies that its overlap with the exact ground state wavefunction
could approach 1 in the limit of large $N$. For instance, the overlap
with a Gaussian of the same width indeed converges to 1 in the
thermodynamic limit.

\begin{centering}
\begin{figure}
\vskip 0 pt
\includegraphics[width=\columnwidth,clip]{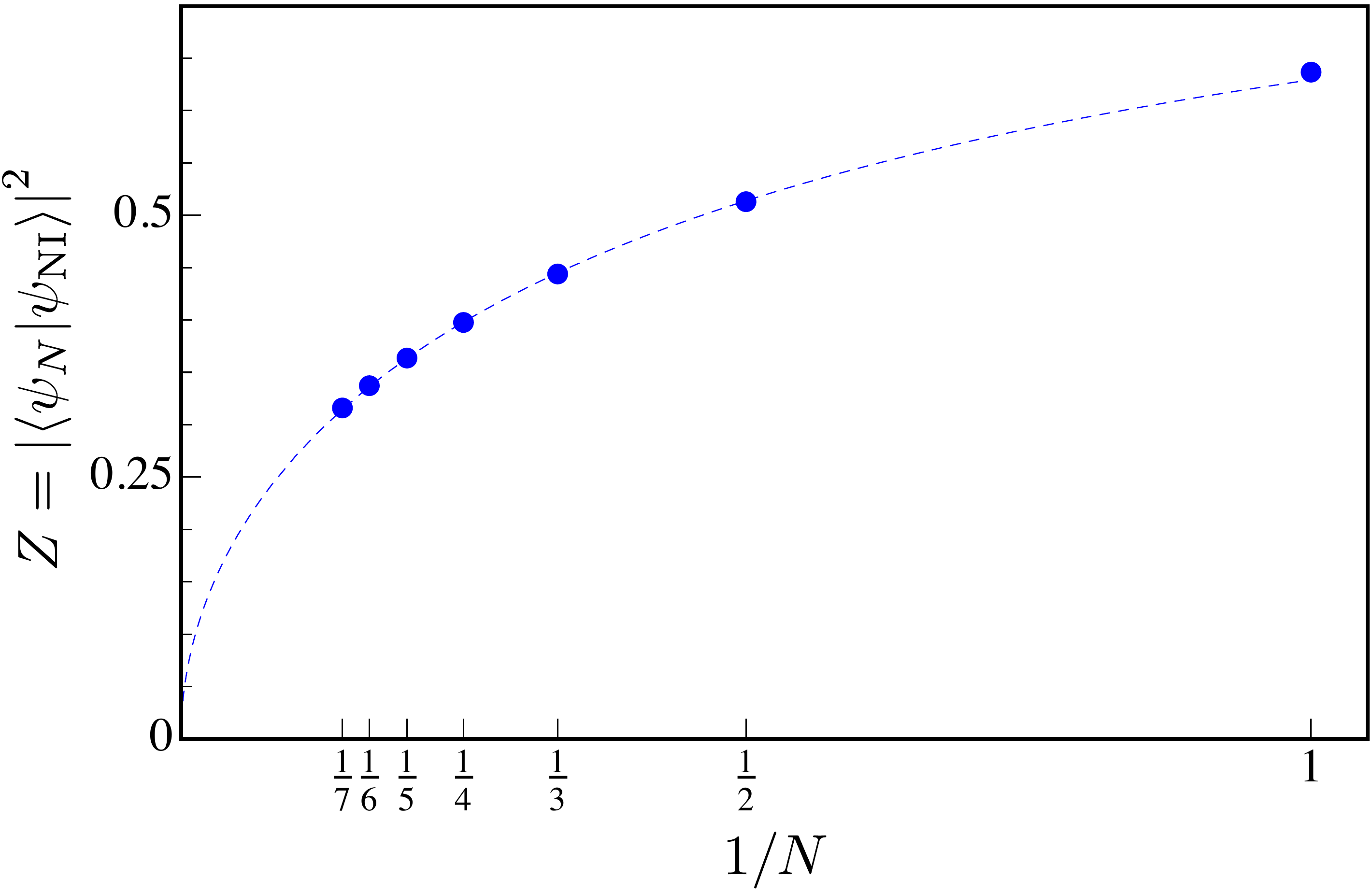}
\caption{ {\bf Emergence of the orthogonality catastrophe.}  Residue
  of the wavefunction $\psi_N$ as a function of $N$. For $N=1$ and
  $N=2$ we find the analytic results $2/\pi$ and $81/(16\pi^2)$,
  respectively. The dashed line is $0.89/\sqrt{N+1}$. We do not show a
  comparison between the residue of the ground state in our
  approximation scheme and that of the exact ground state, as the
  relative error is less than $0.07\%$ for $N\leq7$.}
\label{fig:residue}
\end{figure}
\end{centering}

The small change in the ground-state impurity probability distribution
from the non-interacting to the TG limit appears to suggest that the
wavefunction of the system is only weakly perturbed by infinite
interactions. On the other hand, it is well known that the system
encounters the orthogonality catastrophe in the thermodynamic limit,
where the state of the system has no overlap with the non-interacting
state~\cite{Anderson1967}. To reconcile these points, it is necessary
to consider the impact of the interactions on the majority fermions,
which reshuffle the Fermi sea. This is embedded in the residue
$Z=|\bra{\psi_N}\psi_{\rm \tiny NI}\rangle|^2$, i.e.~the squared
overlap of the ground-state wavefunction with the non-interacting
ground state at $g=0$:
\begin{align}
  \psi_{\rm \tiny NI}(\x) &=\frac{{\cal
      N}_{N-1}}{\pi^{1/4}}\left(\prod_{1\leq i<j\leq
      N}x_{ij}\right)e^{-\sum_{k=0}^Nx_k^2/2} \ . \label{eq:NI}
\end{align}
We compute the residue using a numerical method similar to that
outlined in the Methods, and the result is shown in
Fig.~\ref{fig:residue}. By fitting, we find that the residue decreases
with $N$ as $\sim 1/\sqrt {N+1}$. Intriguingly, the same scaling of
the residue with particle number was predicted for a massive impurity
immersed in a 1D Fermi gas in uniform space \cite{Castella1996}.
\\

\paragraph*{\bf Higher energy manifolds and breathing modes.}
We have demonstrated that our ansatz is extremely accurate for the
$(N+1)$-dimensional ground-state manifold of the impurity problem to
order $1/g$. We now show how it can be extended to states in higher
energy manifolds. It is known that the 2D version of the Hamiltonian
(\ref{eq:H}) is part of a ``spectrum generating'' $SO(2,1)$ algebra
connected with scale transformations
${\mathbf x}\rightarrow{\mathbf x}/\lambda$~\cite{Pitaevskii1997}.  In
1D, this symmetry is broken for a finite interaction strength since
the scaling of the interaction,
$g\delta(x)\rightarrow\lambda g\delta(x)$, is different than in 2D.
The key point, however, is that the $SO(2,1)$ symmetry is recovered in
the TG limit $g\rightarrow\infty$. We can then use the technology
developed for the $SO(2,1)$ symmetry, suitably adapted to the 1D case.
In particular, one can define an operator $\hat B$ so that if
$|n\rangle$ is an eigenstate with energy $E_n$, then
$|n+1\rangle=\hat B^\dagger|n\rangle$ is an eigenstate with energy
$E_{n+1}=E_n+2$ (see Methods). The spectrum in the TG limit therefore
consists of towers of states separated by twice the harmonic potential
frequency, where $\hat B|0\rangle=0$ for the lowest state in each
tower.

Away from the TG limit, each level in these towers is shifted in
energy, and Eq.~\eqref{eq:spectrum} gives the energy shift of the
ground-state manifold to a very good approximation. Each state in the
ground-state manifold represents the lowest state in a separate tower
of modes, and we now use the $SO(2,1)$ algebra to calculate the energy
shift of the excited states in each tower.  Within first order
perturbation theory, the energy shift $\delta E_n$ of the $n$'th
excited mode $|n\rangle$ away from its value in the TG limit is given
by
\begin{align}
  \delta E_n=-\frac1g\frac{\langle0|{\hat B}^n{\cal H}'(\hat
    B^\dagger)^n|0\rangle}{\langle0|\hat B^n(\hat
    B^\dagger)^n|0\rangle}.
\label{Pert}
\end{align}
Note that this result is {\em exact} to order $1/g$. The expectation
values in Eq.~\eqref{Pert} can be calculated using operator algebra
only, once the so-called ``scaling dimension'' $\Delta_{{\cal H}'}$ of
${\cal H}'$ is known. Using Eq.~(\ref{eq:Hp}), we find
 \begin{align}
   \langle\bar \phi_l|{\cal H}'|\bar \phi_n\rangle=\frac 1 {\lambda^3}
   \langle\phi_l| {\cal H}'|\phi_n\rangle,
   \label{ScalingDim}
\end{align}
where
$\bar\phi_n({\mathbf x})=\phi({\mathbf x}/\lambda)/\lambda^{(N+1)/2}$.
It follows that the scaling dimension of ${\cal H}'$ is
$\Delta_{{\cal H}'}=3$ in 1D. One can now use Eq.~(\ref{Pert}) to
express the energy shift of the state $|n\rangle$ as a function of the
energy shift of all lower states in the tower (see Methods).  The
simplest case is the energy shift $\delta E_1$ of the first excited
breathing state $|1\rangle=\hat B^\dagger |0\rangle$, given by
\begin{align}
\delta E_1=\left(1+\frac{3}{4E_0}\right)\delta E_{0},
\label{Lowestbreathing}
\end{align}
where $\delta E_{0}$ is the energy shift of the $N+1$ particle ground
state away from the value $E_0=N(N+2)/2$ in the TG limit. Equation
(\ref{Lowestbreathing}) predicts that the energy shift of the first
excited mode is larger than the shift of the state in the ground-state
manifold. Physically, this means that the excited state energy
approaches its non-interacting value faster than the ground state as
one moves away from the TG limit. Used in combination with Eq.\
(\ref{eq:spectrum}), Eq.\ (\ref{Lowestbreathing}) generalises our
ansatz for the spectrum in the single impurity problem to higher
energy manifolds. However, we emphasize that our results for the
excited manifolds only depend on the energy levels in the ground state
manifold, and are not limited to the impurity problem.
 
We can compare the prediction of Eq.~(\ref{Lowestbreathing}) with the
exact solution to the two-body problem. In 1D, the exact two-body
energies $E$ are determined by the equation
$\Gamma(3/4-E/2)/\Gamma(1/4-E/2)=-g/2\sqrt2$~\cite{Busch1998}. Close
to the TG limit $g\rightarrow \infty$, we have $E=3/2+\delta E_0$ for
a state in the ground-state manifold and $E=7/2+\delta E_1$ for the
first excited state in the tower, with $\delta E_i/E\ll 1$.  Expanding
the $\Gamma$ functions yields $\delta E_1/\delta E_0=3/2$, which is
identical to the result obtained from Eq.~(\ref{Lowestbreathing}) when
using $E_0=3/2$. This demonstrates explicitly that
Eq.~(\ref{Lowestbreathing}), valid for any $N$, recovers the exact
two-body theory close to the TG limit.

For large $N$, it immediately follows from Eq.~\eqref{Lowestbreathing}
that the correction to the energy shift of the first excited manifold
goes as $1/N^2$. Moreover, this holds for any $n\ll N$ (see Methods).
Thus, in the thermodynamic limit we find that the dynamic $SO(2,1)$
symmetry extends to {\em finite} interactions, up to order $1/g$.

The lowest breathing mode frequency has been measured in several
atomic gas experiments~\cite{Kinast2004,Bartenstein2004,Vogt2012}.
The results in this section can therefore be tested experimentally
providing a sensitive probe of interactions in the few-body system.

We note that the approach described in this section is exact to lowest
order in $1/g$, and it is completely general. It would for instance be
interesting to apply it to a system of 1D bosons close to the TG
limit, where the frequency of the lowest breathing mode was recently
calculated using a mapping to an effective fermionic
Hamiltonian~\cite{Zhang2014}.
\\

\paragraph*{\bf Radio-frequency spectroscopy.} 
Our results may be probed directly in cold atomic gases using
radio-frequency (RF) spectroscopy, as already applied in the recent
experiment \cite{Wenz2013}.  Consider a homogeneous RF-probe with
frequency $\omega_{\rm rf}$ which flips the impurity atom from the
hyperfine state $|a\rangle$ to the state $|b\rangle$. Within linear
response, the RF signal is proportional to
\begin{align}
\nn \sum_{i,f}(P_i-P_f) \left|\langle f|{\textstyle
      \int} dx\,\hat\psi_b^\dagger(x)\hat\psi_a(x)|i\rangle\right|^2
  \delta(\omega_{\rm rf}+E_i-E_f),
\end{align}
where $P_i (P_f)$ is the probability of occupation of the initial
$|i\rangle$ (final $|f\rangle$) many-body state, and $\hat\psi_\sigma$
is the field operator for the hyperfine state $|\sigma\rangle$.

Assume that the system initially is in a definite state $|i\rangle$
and that all final states are empty. There are two kinds of RF
spectroscopy. In \emph{direct} RF-spectroscopy, $a=\downarrow$ and the
impurity atom interacts with the $\uparrow$ atoms in the initial
state, which belongs to the interacting many-body ground-state
manifold, whereas the final hyperfine state $|b\rangle$ of the
impurity atom does not interact with the majority atoms. There will
then be a peak at $\omega_{\rm rf}=-E_0$ in the RF spectrum in the TG
limit, and the reduction of the height of the peak from its
non-interacting value gives the quasiparticle residue of the initial
state. There will also be peaks at $\omega_{\rm rf}=-E_0+2n$ with
$n=1, 2,\ldots$ as the initial interacting wavefunction has components
in excited non-interacting states with the same parity. The shift of
the peak position away from $\omega_{\rm rf}=-E_0$ gives the energy
shift of the many-body ground state when $1/g>0$.  In \emph{inverse}
RF spectroscopy, the initial state $|a\rangle$ of the impurity atom
does not interact with the majority atoms whereas the final state does
with $b=\downarrow$~\cite{Kohstall2011}. There will then be a peak at
$\omega_{\rm rf}=E_0$ in the RF spectrum in the TG limit and the shift
in position when $1/g>0$ again gives the many-body energy shift
directly.  The reduction of the height of the peak from its
non-interacting value gives the quasiparticle residue. There will also
be RF peaks at higher frequencies corresponding to flipping into the
excited interacting states.
\\

\section*{Discussion}
In this work, we investigated in detail the properties of a single
impurity immersed in a Fermi sea of $N$ majority particles near the TG
limit.  By comparing with exact numerical results, we have
demonstrated the impressive accuracy of our strong-coupling ansatz for
arbitrary $N$.  We have furthermore identified the effective
Heisenberg Hamiltonian within which our ansatz is exact, and this has
allowed us to evaluate analytically the entire ground-state manifold,
yielding the discrete Chebyshev polynomials.  In particular, the
ground-state wavefunction from our ansatz at strong repulsion is a
sign-alternating Pascal's triangle in the spin chain. Since its
overlap with the exact ground-state wavefunction extrapolates to a
value $\sim0.9999$ for $N\to\infty$, we believe that Eq.~\eqref{eq:GS}
is essentially indistinguishable from the result of numerically exact
approaches.

In addition to the static properties considered here, our ansatz
provides a framework for investigating impurity dynamics in a harmonic
potential, since we have determined the entire spectrum of the
ground-state manifold and associated excited states related via a
scale transformation.  The impurity dynamics in 1D gases have recently
been investigated theoretically in Refs.~\cite{Mathy2012,Kantian2014},
and experimentally in Refs.~\cite{Catani2012,Fukuhara2013}.

Our results also extend far beyond the single-impurity problem since
the effective Heisenberg Hamiltonian \eqref{eq:lattice} accurately
describes any number of $\up,\,\down$ particles in the strongly
coupled regime, as explicitly demonstrated for $N_\up=N_\down=2$. For
larger $N_\up$ and $N_\down$, where the number of states grows
dramatically, our Hamiltonian can be tackled with numerical tools
developed for lattice systems, such as the density matrix
renormalization group \cite{White1992}, or matrix product states
\cite{Affleck1988}.  Extending our approach beyond the two-component
Fermi gas to other quantum mixtures would also enable us to address
open problems in the context of quantum magnetism, such as the nature
of correlations and dynamical quantum phase transitions.  In
particular, it would be interesting to investigate whether our ansatz
may be extended to harmonically confined $N$-component fermions, a
scenario which has relevance to $SU(N)$ magnetism~\cite{Gorshkov2010}
and which has recently been experimentally realized \cite{Pagano2014}.

Finally, the exceptional accuracy of our simple ansatz and the
suggestive form of the impurity spectrum $E-E_0\propto l(l+1)/g$,
lead us to speculate that our results are the manifestation of a
hidden approximate symmetry.  This raises the tantalizing possibility
that the 1D Fermi gas in a harmonic potential becomes near integrable
in the strongly interacting limit.
\\

%%%%%%%%%%%%%%%%%%%%%%%%%%%%%%%%%%%%%%%%%%%%%%%%%

\section*{Methods}

\paragraph*{\bf Manipulations of the basis functions.}
In order to calculate the overlaps of the basis functions $\phi_l$ it
is useful to introduce an alternative formulation of the
problem. First we note that, for a given ordering of particles, all
basis functions $\phi_l$ (and consequently any superposition of these)
are proportional to $\psi_0$. We may then define the complete and
orthonormal set of basis states:
\begin{align}
  \langle \x\ket{\down_i} = \sqrt{N+1} \
  \psi_0(\x) \Theta\left(\{ x_j
    \}_i < x_0< \{ x_j \}_{N-i} \right),
\label{eq:downi}
\end{align}
where $1 \leq j \leq N$. $\{ x_j \}_i$ corresponds to any set of $i$
spin-$\up$ particles and the step function $\Theta$ is 1 if precisely
$i$ of the $N$ majority particles are to the left of the impurity, and
zero otherwise. Clearly the states described by Eq.~\eqref{eq:downi}
do not overlap. To see that they are properly normalized, consider
\begin{align}
  \langle\down_i \ket{\down_i} &= (N+1) \!  \int \!  d\x \
  |\psi_0(\x)|^2 \Theta\left(\{ x_j
    \}_i <\! x_0\!< \{ x_j \}_{N-i} \right) \nn \\
  &=(N+1) \frac{N!}{(N+1)!}=1.
\label{eq:downi2}
\end{align}
Here we used the fact that the integral over the (normalized)
ferromagnetic state $\psi_0$ does not depend on the ordering of
particles. Thus, restricting the integral to a particular ordering,
the result is $1/(N+1)!$. Now, if $i$ particles are to the left of the
impurity, there are ${{N}\choose{i}}$ ways of choosing these, with
$i!\times (N-i)!$ ways of ordering those to the left and right of the
impurity.  Gathering the terms, the basis states
$\{\left|\down_i\right>\}$ are thus seen to form an orthonormal basis.

Recall now the definition of the basis states $\phi_l$:
\begin{align}
  \phi_l&=\psi_0(\x)\sum_{1\leq i_1<\cdots<i_l\leq N}s_{i_1}\cdots
  s_{i_l}, \hspace{5mm} 1\leq l\leq N,
\end{align}
with $s_j\equiv \mbox{sign}(x_j-x_0)$. Assuming there are exactly $i$
majority particles to the left of the impurity, we may evaluate the
sum of sign functions:
\begin{align}
\sum_{1\leq i_1<\cdots<i_l\leq N}s_{i_1}\cdots
s_{i_l}=\sum_{k=0}^l(-1)^k
\binomial{i}{k}\binomial{N-i}{l-k}.
\label{eq:signsum}
\end{align}
This simply counts how many combinations exist with $k$ [$l-k$]
majority particles involved in the sign functions out of the $i$
[$N-i$] majority particles located to the left [right] of the
impurity, respectively. Thus we arrive at the projection of the two
sets of basis states:
\begin{align}
  \bra{\down_i}\phi_l\rangle=
  \frac1{\sqrt{N+1}}\sum_{k=0}^l(-1)^k\binomial{i}{k}
  \binomial{N-i}{l-k}.
\label{eq:basischangeapp}
\end{align}
The prefactor arises from the same arguments that led to
Eq.~\eqref{eq:downi}.

The inner products $\Phi_{ln}=\bra{\phi_l}\phi_n\rangle$ then simply
follow from Eq.~\eqref{eq:basischangeapp},
\begin{align}
\Phi_{ln}=\sum_{i=0}^N \bra{\phi_l}\down_i
&\rangle\langle\down_i\ket{\phi_n} \nn\\
=\frac1{N+1} \sum_{i=0}^N 
&\left[ \sum_{k_1=0}^l (-1)^{k_1} \binomial{i}{k_1} \binomial{N-i}{l-k_1} \right]
\nn \\ \times &
\left[ \sum_{k_2=0}^n (-1)^{k_2} \binomial{i}{k_2} \binomial{N-i}{n-k_2}
\right].\nn
\end{align}
We also note that the matrix $\Phi$ is bisymmetric,
i.e.~$\Phi_{ln}=\Phi_{nl}=\Phi_{N-l,N-n}$.
\\

\paragraph*{\bf Numerical evaluation of ${\cal H}'$.} 
The matrix elements of ${\cal H}'$ can be evaluated as
\begin{align} {\cal H}'_{ln} & =\bra{\phi_l} {\cal H}' \ket{\phi_n} =
  \lim_{g \to \infty} g^2 \sum_{i=1}^N \int d\x \ \delta(x_{i0})
  \phi_l \phi_n
  \nn \\
  & = \sum_{i=1}^N \int d\x \ \delta(x_{i0}) \left. \frac{\partial
      \phi_l }{\partial x_{i0}} \right|^+_- \left. \frac{\partial
      \phi_n }{\partial x_{i0}} \right|^+_-,
\label{eq:Hp2}
\end{align}
where we have used the fact that
$g\phi_l(x_{i0}=0)= \lim_{x_{i0}\to0^+} \frac{\partial \phi_l
}{\partial x_{i0}}(x_{i0})-\lim_{x_{i0}\to0^-} \frac{\partial \phi_l
}{\partial x_{i0}}(x_{i0})\equiv \frac{\partial \phi_l }{\partial
  x_{i0}}|_-^+$.
This quantity can be non-zero due to the presence of cusps in the
basis functions. Note that
$\frac{\partial \phi_0}{\partial x_{i0}}|_-^+ = 0$ for the
ferromagnetic state and therefore
${\cal H}'_{0n} = {\cal H}'_{n0} = 0$. This implies that $\phi_0$ is
an eigenstate with $\mathcal{C}=0$, as expected.

We now show that the $(N+1)$-dimensional integrals appearing in the
matrix elements of $\mathcal{H}'$, Eq.~\eqref{eq:Hp2}, may
conveniently be expressed in terms of a Taylor expansion of a suitable
function.  We begin by noting that the bracketed term of the
wavefunction
\begin{align}
  \psi_0(\x) &=\NN\left(\prod_{0\leq i<j\leq
      N}x_{ij}\right)e^{-\sum_{k=0}^Nx_k^2/2} ,
\end{align}
only depends on the relative coordinates. It is then convenient to
introduce the center-of-mass coordinate
$x_{\rm cm}\equiv\sum_{i=0}^N x_i/(N+1)$, and relative coordinates
$y_0 = (N+1)(x_0-x_{\rm cm})$ and $y_i= x_{i0}$ $(1\leq i\leq N)$.
The constraint $\sum_{i=0}^N x_i-(N+1)x_{\rm cm}=\sum_{i=0}^Ny_i=0$
may be enforced through the use of a $\delta$-function, which in turn
may be converted into an extra integral, yielding
\begin{equation*}
  \int d\x \ (\cdot)=\int\frac{dk}{2\pi}dx_{\rm cm}  \,
  d\mathbf{y}\,e^{ik\sum_{j=0}^N y_{j}} \ (\cdot) 
\end{equation*}
This allows us to decouple the center-of-mass coordinate. The
integrand of Eq.\ \eqref{eq:Hp2} contains a factor of
$e^{-\sum_{i=0}^Nx_i^2}$ (following from the form of $\psi_0$), which
may be written as
$e^{-\sum_{i=1}^Ny_{i}^2+y_0^2/(N+1)-(N+1)x_{\rm cm}^2}$. As this is
the only term in which the center-of-mass coordinate appears in
\eqref{eq:Hp2}, $x_{\rm cm}$ may be integrated out to give
$\sqrt{\frac\pi{N+1}}$.

Next consider the part of the integral involving both $k$ and $y_0$:
\begin{align*}
  \int \frac{dk}{2\pi} dy_0\, (\cdot) e^{y_0^2/(N+1)}e^{iky_0} = &\int
  \frac{dk}{2\pi} dy_0\, (\cdot)  e^{-\nabla_k^2/(N+1)}e^{iky_0}\\
  = &\int dk\, \delta(k)\, e^{-\nabla_k^2/(N+1)} (\cdot) ,
\end{align*}
where in the last step we have integrated by parts. Collecting all
factors, we get an expression where the integrals over the $y_i$ have
been decoupled:
\begin{align} {\cal H}'_{ln} = & \sum_{i=1}^N \int d\x \
  \delta(y_{i}) \left. \frac{\partial \phi_l }{\partial y_{i}}
  \right|^+_-
  \left. \frac{\partial \phi_n }{\partial y_{i}} \right|^+_- \nn \\
  = & 4N{\cal N}_N^2\sqrt{\frac{\pi}{N+1}}\int dk \, \delta(k)
  e^{-\nabla_k^2/(N+1)} \nn \\ & \times \left[ \int
    d^{N-1}y\,e^{\sum_{j=1}^{N-1}(iky_j-y_j^2)}
    S_{l-1}^{(N-1)}S_{n-1}^{(N-1)} \right. \nn \\ & \left. \hspace{10mm}\times\prod_{1\leq i_1<i_2<
      N}(y_{i_1}-y_{i_2})^2\prod_{1\leq m< N}y_m^2\right],
\end{align}
where
$S_l^{(N)}\equiv\sum_{1\leq i_1<\cdots<i_l\leq N}s_{i_1}\cdots
s_{i_l}\prod_{i=1}^Ny_i$.
We have made use of the fact that the integral is independent of $i$
so that we can differentiate with respect to just $y_N$ and multiply
by $N$. We also used the relation
$\left. \frac{\partial S_l^{(N)}}{\partial y_N}
\right|^+_-=2S_{l-1}^{(N-1)}$.
Denoting the quantity in square parentheses by $h_{ln}(k)$, and
introducing its Taylor expansion around $k=0$,
$h_{ln}(k)=\sum_{m=0}^\infty h_{ln}^{(m)} k^m$, we see that the
desired matrix element is converted into a quickly convergent series,
${\cal H}'_{ln}=4N{\cal
  N}_N^2\sqrt{\frac{\pi}{N+1}}\sum_{m=0}^\infty\left(\frac {-1} {N+1}
\right)^m\frac{(2m)!}{m!}h_{ln}^{(2m)}$.

For the example of $N=3$, the function appearing in square parentheses
for the matrix element ${\cal H}'_{11}$ is
$h_{11}(k)=\int dy_1\,dy_2\,e^{ik(y_1+y_2)-y_1^2-y_2^2}y_1^4
y_2^4(y_1-y_2)^2$,
while for the matrix element ${\cal H}'_{31}$, it is
$h_{31}(k)=\int dy_1\,dy_2\,e^{ik(y_1+y_2)-y_1^2-y_2^2}|y_1y_2|y_1^3
y_2^3(y_1-y_2)^2$.
In both cases we see how the integrals over the relative coordinates
between the impurity and the majority particles separate. It is also
easy to see that $h_{11}(k)=h_{33}(k)$, as expected.
\\

\paragraph*{\bf Analytic solution for $\bm{N=3}$.} 
We start by considering the explicit solution of the Schr{\"o}dinger
equation perturbatively for $1/|g|\ll1$. First, we note that the
matrix elements of ${\cal H}'$ can be written as
\begin{align} 
\notag {\cal H}'_{ln} &= 4 N \sum_{i=0}^{N-1} {\cal I}_i
  \left[ \sum_{k_1=0}^{l-1} (-1)^{k_1} \binomial{i}{k_1}
    \binomial{N-1-i}{l-1-k_1} \right] \\ \notag &\times
  {{N-1}\choose{i}} \left[ \sum_{k_2=0}^{n-1} (-1)^{k_2}
    \binomial{i}{k_2} \binomial{N-1-i}{n-1-k_2} \right].
\end{align}
where we have the integral
\begin{align}\notag
{\cal I}_i & = 
\int^\infty_{-\infty} dx_0 \int^{x_0}_{-\infty} d^ix_j
\int^{\infty}_{x_0} d^{N-1-i}x_k
\left.\left(\frac{\partial\psi_0}{\partial x_{N0}}\right)^2
\right|_{x_{N0}=0}
\end{align} 
i.e., we integrate over $i$ of the majority particles to the left of
the impurity, and $N-i-1$ to the right.  Note that the integral only
couples basis states with the same parity. One can also show that
\begin{align} \notag
\bra{\phi_{N-l+1}} {\cal H}' \ket{\phi_{N-n+1}} = \bra{\phi_l} {\cal H}' \ket{\phi_n}.
\end{align}

For $N=3$, it is possible to evaluate ${\cal I}_i$ analytically, thus
giving
\begin{align} \notag {\cal H}'_{11} & = {\cal H}'_{33} = \frac{475}{32
    \sqrt{2\pi}}, \\ \notag
  {\cal H}'_{13} & = {\cal H}'_{31} = \frac{1425 \sqrt{2} \pi -56 -2850 \sqrt{2} \tan^{-1}\left(2 \sqrt{2}\right)}{192 \pi ^{3/2}}, \\
  {\cal H}'_{22} & = \frac{1425 \sqrt{2} \pi -28-1425 \sqrt{2} \tan
    ^{-1}\left(2 \sqrt{2}\right)}{48 \pi ^{3/2}}.
\end{align}
Combining this with the matrix of inner products $\Phi$ then allows us
to solve the eigenvalue equation and determine $\mathcal{C}$ and
$\psi$ analytically.  The resulting expressions are rather cumbersome
and yield the numerical values shown in the main text.
\\

\paragraph*{\bf \bm{$SO(2,1)$} symmetry and excited states.} 
We define the scaling operator $\hat S(\lambda)$ as 
\begin{align}
  \hat S(\lambda)\psi({\mathbf x})=\frac{\psi({\mathbf
      x}/\lambda)}{\lambda^{(N+1)/2}}.
\end{align}
It can be written as
\cite{Nishida2007,Werner2006}
\begin{align}
  \hat S(\lambda)=e^{-i\ln\lambda \hat D}\hspace{0.5cm}\text{ with
  }\hspace{0.5cm}\hat D=\frac 1 {2i}\sum_j(x_j\partial_j+\partial_jx_j).
\end{align}
The raising and lowering operators can then be defined
as~\cite{Moroz2012}
\begin{align}
  \hat B=\hat L-\frac{\hat Q^2}{N+1}\hspace{0.5cm}\text{ with }
  \hspace{0.5cm}\hat L=\frac 12({\cal H}-\hat X^2-i\hat D),\nn
\end{align}
for the $N+1$ particle state. Here $\hat X^2=\sum_ix_i^2$ is twice the
harmonic potential, and $\hat Q=(\hat P+i\hat K)/\sqrt 2$ with $\hat
P=\sum_j \hat p_j$ the total momentum and $\hat K=\sum_jx_j$ the
center-of-mass coordinate. One can show that, in the TG
  limit, $[{\cal H},\hat B^\dagger]=2\hat B^\dagger$ from which it
follows that if $|n\rangle$ is an eigenstate with energy $E_n$, then
$|n+1\rangle=\hat B^\dagger|n\rangle$ is an eigenstate with energy
$E_{n+1}=E_n+2$. The operator $\hat B^\dagger$ excites breathing
modes, and the spectrum in the TG limit consists of towers of states
separated by twice the harmonic potential frequency, where $\hat
B|0\rangle=0$ for the lowest state in each tower.

To calculate the scaling dimension, we take the derivative of
Eq.~(\ref{ScalingDim}) with respect to $\lambda$ setting $\lambda=1$,
from which it follows that the scaling dimension is
$\Delta_{{\cal H}'}=3$. The calculation of the energy shift,
Eq.~(\ref{Pert}), is now rather long and cumbersome but
straightforward since all necessary commutators are known
\cite{Moroz2012}. We obtain
\begin{align}
  \delta E_n-\delta E_{n-1}=\frac{3}{4R_{n-1}}\delta
  E_{n-1}+\frac{R_{n-2}}{R_{n-1}}(\delta E_{n-1}-\delta E_{n-2}),
\label{BreathingShift}
\end{align}
where we have defined
$R_n=\sum_{j=0}^nE_j=\sum_{j=0}^n(E_0+2j)=(n+1)(E_0+n)$. Here $E_j$ is
the internal energy of the $j$'th state, i.e the energy minus the zero
point energy of the center-of-mass. The last term in
Eq.~(\ref{BreathingShift}) is zero for $n=1$. Equation
(\ref{BreathingShift}) gives the shift $\delta E_n$ of the
energy of the state $|n\rangle$ away from its value in the TG limit in
terms of the energy shifts of the lower modes.

\bibliography{gramschmidt}

%%%%%%%%%%%%%%%%%%%%%%%%%%%%%%%%%%%%%%%%%%%%%%%%%
{\bf Acknowledgments:}
We gratefully acknowledge enlightening discussions with Selim Jochim
and Gerhard Z{\"u}rn on spin-polarized 1D experiments, and feedback on
this work from Gareth Conduit, Frank Deuretzbacher, Vudtiwat
Ngampruetikorn, David Petrosyan, Luis Santos, Richard Schmidt, Vijay
Shenoy, Artem Volosniev, Zhenhua Yu, and Nikolaj Zinner. We thank
Giuseppe Fedele for useful correspondence on discrete polynomials.
{\bf Funding:} P.M. acknowledges support from ERC AdG OSYRIS, EU EQuaM
and IP SIQS, Plan Nacional FOQUS and the Ram\'on y Cajal programme,
while M.M.P. acknowledges support from the EPSRC under Grant No.\
EP/H00369X/2. GMB would like to acknowledge the support of the
Hartmann Foundation via grant A21352 and the Villum Foundation via
grant VKR023163.  J.L., P.M., and M.M.P. also wish to thank the
Institute for Nuclear Theory at the University of Washington where
this work was initiated. Finally, we wish to thank the ESF POLATOM
network for financial support.

\end{document}